\RequirePackage{fix-cm}

\documentclass[twocolumn,epjc3]{svjour3}
\smartqed

\RequirePackage{graphicx}

\usepackage{xcolor}
\usepackage{soul}
\usepackage{caption}
\usepackage{booktabs}
\usepackage{upgreek}
\usepackage{mdframed}
\usepackage{acro}
\usepackage{isotope}
\usepackage{orcidlink}
\usepackage[T1]{fontenc}
\usepackage[utf8]{inputenc}
\usepackage{hyperref}
\usepackage{amsmath}

\hypersetup{
    colorlinks=true,
    linkcolor=blue,
    citecolor=blue,
    urlcolor=blue
}

\DeclareAcronym{PE}{  short=PE,  long=photoelectron,}
\DeclareAcronym{SPE}{  short=SPE,  long=single photoelectron,}
\DeclareAcronym{TPC}{  short=TPC,  long=Time Projection Chamber,}
\DeclareAcronym{LNGS}{ short=LNGS, long=Laboratori Nazionali del Gran Sasso,}
\DeclareAcronym{WIMP}{ short=WIMP, long=Weakly Interacting Massive Particle,}
\DeclareAcronym{DM}{   short=DM,   long=dark matter,}
\DeclareAcronym{HV}{   short=HV,   long=high voltage,}
\DeclareAcronym{GN2}{  short=GN2,  long=gaseous nitrogen,}
\DeclareAcronym{GXe}{  short=GXe,  long=gaseous xenon,}
\DeclareAcronym{LN2}{  short=LN2,  long=liquid nitrogen,}
\DeclareAcronym{LXe}{  short=LXe,  long=liquid xenon,}
\DeclareAcronym{Xe}{  short=Xe,  long=xenon,}
\DeclareAcronym{PMT}{  short=PMT,  long=photomultiplier tube,}
\DeclareAcronym{ER}{   short=ER,   long=electronic recoil,}
\DeclareAcronym{NR}{   short=NR,   long=nuclear recoil,}
\DeclareAcronym{SM}{   short=SM,   long=Standard Model,}
\DeclareAcronym{GERDA}{short=GERDA,long=GERmanium Detector Array,}
\DeclareAcronym{NEMO}{ short=NEMO, long=Neutrino Ettore Majorana Observatory,}
\DeclareAcronym{KamLAND-ZEN}{ short=KamLAND-ZEN, 
  long=Kamioka Liquid Scintillator Antineutrino Detector--Zero Neutrino Double Beta Decay Experiment,}
\DeclareAcronym{CES}{  short=CES,  long=combined energy scale,}
\DeclareAcronym{NEXT}{ short=NEXT, long=Neutrino Experiment with a Xenon TPC,}
\DeclareAcronym{EXO-200}{ short=EXO-200, long=Enriched Xenon Observatory,}
\DeclareAcronym{2nu2b}{ short=$2\upnu\upbeta\upbeta$, long=two-neutrino double beta decay,}
\DeclareAcronym{0nu2b}{ short=$0\upnu\upbeta\upbeta$, long=neutrinoless double beta decay,}
\DeclareAcronym{HPXe}{ short=HPXe, long=high-pressure gaseous xenon,}
\DeclareAcronym{PSD}{ short=PSD, long=pulse shape discrimination,}
\DeclareAcronym{RoI}{ short=RoI, long=region of interest,}
\DeclareAcronym{PTFE}{ short=PTFE, long=polytetrafluoroethylene,}
\DeclareAcronym{DEC}{ short=DEC, long=double electron capture,}
\DeclareAcronym{FV}{ short=FV, long=fiducial volume,}
\DeclareAcronym{CEVNS}{ short=CE$\upnu$NS, long=coherent elastic neutrino-nucleus scattering,}
\DeclareAcronym{LCE}{ short=LCE, long=light collection efficiency,}
\DeclareAcronym{NEST}{ short=NEST, long=Noble Element Simulation Technique,}
\DeclareAcronym{CL}{ short=C.L., long=confidence limit,}

\newcommand{\Xenat}{\ensuremath{^{\mathrm{nat}}\mathrm{Xe}}\,}
\newcommand{\qvalue}{$Q_{\upbeta\upbeta}$}

\journalname{Eur. Phys. J. C}

\begin{document}
\title{HERETIX: A Hermetic, Enriched, Rare-Event Time Projection Chamber in Xenon}

\author{
L.~Althueser\thanksref{1}\,\orcidlink{0000-0002-5468-4298}
\and
N.~Hargittai\thanksref{2}\,\orcidlink{0009-0007-8906-2011}
\and
J.~Pienaar\thanksref{2,3}\,\orcidlink{0000-0001-5830-5454}
\and
R.~Braun\thanksref{1}\,\orcidlink{0009-0007-0706-3054}
\and
H.~Landsman\thanksref{2}\,\orcidlink{0000-0002-7570-5238}
\and
C.~Weinheimer\thanksref{1}\,\orcidlink{0000-0002-4083-9068}
\and
R.~Budnik\thanksref{2}\,\orcidlink{0000-0002-1963-9408}
}

\institute{Institut f\"ur Kernphysik, University of M\"unster, 48149 M\"unster, Germany \label{1}
\and
Department of Particle Physics and Astrophysics,
Weizmann Institute of Science,
Rehovot 7610001, Israel \label{2}
\and
Department of Physics Core Facilities, Weizmann Institute of Science, Rehovot 7610001, Israel \label{3}
}

\date{Received: date / Accepted: date}

\maketitle

\begin{abstract}
Xenon-based time projection chambers have established themselves as one of the most powerful technologies for rare-event searches. HERETIX is a proposed multi-tonne liquid xenon observatory featuring two nested time projection chambers that enable the simultaneous optimisation of searches for weakly interacting massive particles and \ac{0nu2b} of \isotope[136]{Xe}. A hermetically sealed sapphire vessel containing xenon enriched to 90\% \isotope[136]{Xe} forms the inner detector, providing an ultra-low-background environment for \ac{0nu2b} searches. Monte Carlo studies indicate that material-induced backgrounds can be effectively eliminated, yielding a projected \ac{0nu2b} half-life sensitivity of $3.2 \times 10^{28} \, \mathrm{years}$ at 90\% confidence level after a 10-year exposure, while the surrounding xenon volume, depleted in \isotope[136]{Xe}, preserves the excellent dark matter sensitivity of large liquid xenon detectors. HERETIX therefore offers a unified experimental approach capable of delivering leading sensitivity to two of the most compelling questions in fundamental physics.
\end{abstract}

\section{Introduction}
\label{sec:intro}

Rare-event searches play a central role in the effort to uncover new physics beyond the Standard Model. Among these, two stand out for their potential impact: the search for \ac{DM} particles and the search for \ac{0nu2b} decay \cite{MarrodanUndagoitia:2015veg,Roszkowski:2017nbc,Dolinski:2019nrj,RevModPhys.95.025002}. Although motivated by distinct questions of cosmology, such as dark matter and the baryon asymmetry of the universe, and particle physics, such as the nature of neutrinos and physics beyond the Standard Model, both rely on large-scale, underground, ultra-sensitive detectors capable of identifying rare interactions, which require mitigation of backgrounds to extremely low levels~\cite{Misiaszek:2023sxe,Miramonti:2025ges}. These experiments push the limits of detector design, material radio-purity, and statistical analysis, yet despite their different scientific goals, they share many experimental challenges and technologies, offering opportunities for joint developments and coordinated design strategies.

\acp{TPC} which use Xenon as the active target have emerged as a leading technology in both fields~\cite{XENON:2024wpa,LZ-instrument,Auger:2012gs,Gomez-Cadenas:2014dxa}, offering scalability, radio-purity, and the ability to combine calorimetry with spatial reconstruction. However, the distinct objectives of each science case have led to different design choices optimised for their respective unique constraints. \ac{DM} searches in \ac{LXe} \acp{TPC} are best suited to detect \acp{WIMP} with masses of $\mathcal{O}$(10-100\,GeV/c$^{2}$), and thus look for \ac{NR} signals in the $\mathcal{O}$(10\,keV) energy range~\cite{Lewin1996}. This requires low energy thresholds and excellent discrimination against \ac{ER} interactions~\cite{LUX:2020car}, which are the dominant background in the energy \ac{RoI}. XENONnT~\cite{XENON:2024wpa}, LZ \cite{Akerib2024LUXZEPLINDevelopments} and PandaX-4T~\cite{Bo2025PandaX4T154TonneYear} achieve this by utilising a dual-phase TPC design with a \ac{GXe} region at the top. This configuration takes advantage of both the prompt scintillation light produced by a particle interaction in the \ac{LXe} and a delayed charge signal extracted into the \ac{GXe} phase, the latter of which can be detected with nearly 100\% efficiency down to a single extracted electron~\cite{XENON:2024mlv,LUX:2017bef}. The delayed coincidence of the two signals provides a three-dimensional position reconstruction of each event, enabling the suppression of surface and material backgrounds in the energy \ac{RoI} as well as discrimination between \ac{NR} and \ac{ER} interactions.

Xenon contains \isotope[136]{Xe} (natural abundance 8.86\%~\cite{Meija2016AtomicWeights}) which decays via a double $\upbeta$-decay with a Q-value of 2458\,keV~\cite{PhysRevLett.98.053003}, thus motivating experimental campaigns aimed at detecting the \ac{0nu2b} decay process, such as EXO-200~\cite{Anton2019EXO200CompleteDataset}, NEXT~\cite{NEXT:2025yqw} and the planned nEXO~\cite{nEXO:2021ujk}. The Q-value is large relative to many natural radioactive decays, and the energy deposition is expected to be reconstructed as a single interaction due to the short mean free path of electrons in \ac{LXe}. The most significant experimental backgrounds originate from the 2448\,keV $\gamma$-ray emitted during the $\upbeta$-decay of \isotope[214]{Bi}, and the 2615\,keV $\gamma$-ray from \isotope[208]{Tl}. Therefore, \ac{0nu2b} searches with Xenon targets must minimise these experimental backgrounds from contamination of detector materials with \isotope[238]{U} and \isotope[232]{Th} and their decay products which include \isotope[214]{Bi} and \isotope[208]{Tl}, respectively. Typically these backgrounds are suppressed by their spatial distribution. Additionally, a high precision energy resolution well below 1\% is required to distinguish the \isotope[214]{Bi} peak from a potential \ac{0nu2b} signal, as the expected energy separation is only 10\,keV. The signal-to-background ratio for \ac{0nu2b} searches in \ac{LXe} detectors is further enhanced by using isotopically enriched target volumes with \isotope[136]{Xe} concentrations of up to 90\%~\cite{Anton2019EXO200CompleteDataset,NEXT:2025yqw,nEXO:2021ujk}.

To surpass the sensitivity of searches for \ac{DM} particles or \ac{0nu2b} decays beyond the current experimental limits requires significantly greater xenon target masses. Thus it is natural to ask whether the two rare-event searches can meaningfully be pursued in a single multi-tonne detector design. A unified, xenon-based \ac{TPC} design could serve both purposes, maximising physics output from shared xenon handling infrastructure, target mass and shielding. In this work we present HERETIX, a design solution in which both are combined in synergy as one detector. We describe the concept of HERETIX, a hermetically sealed inner \ac{TPC} containing xenon isotopically enriched with \isotope[136]{Xe} within a larger \ac{TPC}. We describe its implementation, estimated background and detector responses in Section~\ref{sec:backgrounds}. As shown in Section~\ref{sec:physics}, our HERETIX proposal will reach world leading sensitivities for the half-life of the \ac{0nu2b}-decay of \isotope[136]{Xe}, while incurring only a modest penalty on the required exposure to achieve the planned \ac{WIMP} sensitivity of the nominal XLZD proposal. Finally, we discuss the specific instrumental and design challenges associated with integrating the requirements of both physics searches into a single detector in Section~\ref{sec:detector}, and consider the impact on the projected sensitivities of some modifications of the nominal HERETIX proposal.\\

\section{HERETIX: A Dual-purpose Experiment}\label{sec:backgrounds}

Focus is shifting towards design decisions required to realise proposals for next-generation \ac{LXe}-based experiments that can achieve sensitivities well into the ``neutrino fog''~\cite{Carew:2023qrj}, such as the XLZD~\cite{XLZD:2024nsu} rare-event observatory for \ac{DM} and neutrino physics. The baseline proposal is a 60\,tonne \Xenat target mass dual-phase \ac{TPC} which is expected to achieve a 90\% exclusion sensitivity of about 1$\times10^{-49}$\,cm$^2$ for 40\,GeV/c$^{2}$ mass \ac{WIMP}s after a 1000\,t$\cdot$y exposure. Additionally, its sensitivity to \ac{0nu2b}-decay half-lives is expected to be up to $6.09\times10^{27}\,\mathrm{years}$~\cite{XLZD:Aalbers2025,XLZD:2024nsu}. Proposals for next generation \ac{0nu2b} detectors, such as nEXO~\cite{nEXO:2021ujk} have been put forward which envisions a 5\,tonne \ac{LXe} target enriched to 90\% \isotope[136]{Xe}. The proposal would have a 3$\sigma$ discovery potential for half-lives up to 7.4$\times10^{27}$\,years.

\begin{figure}[tbp]
	\centering
	\includegraphics[width=0.38\textwidth]{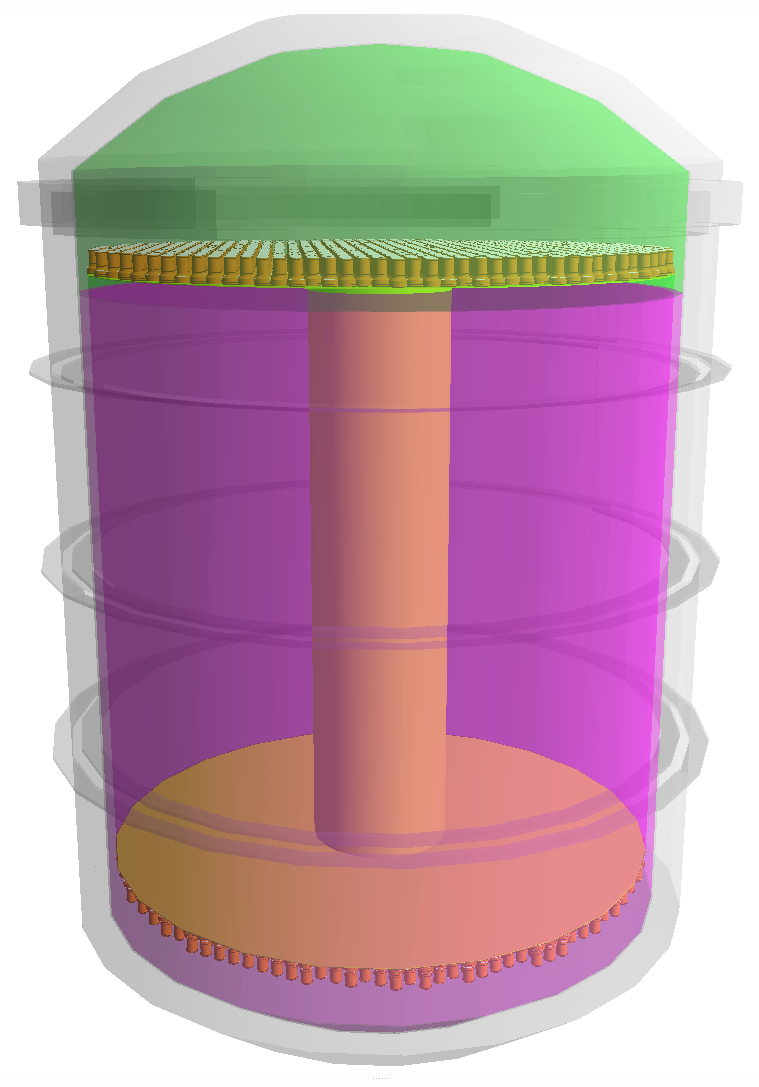}
	\caption{Visualisation of HERETIX as implemented in GEANT4. The detector is contained within a double walled stainless steel cryostat (grey). The active \ac{LXe} target (purple) and \ac{GXe} (green) are situated inside a \ac{PTFE} cylinder. At the center, an inner transparent sapphire cylinder (orange) contains enriched \ac{LXe}. The active region is end-capped by \acp{PMT}.} 
    	\label{fig:G4_vis}
\end{figure}

Realisation of both detectors in parallel would need well in excess of 150\,tonnes of xenon when taking into consideration enrichment requirements, as well as the need for \ac{LXe} mass outside the active volume and within cryogenic, purification and recirculation volumes. We therefore propose a dual-purpose approach which relies on approximately 75\,tonnes of \ac{LXe}, split between a 5\,tonne volume of \ac{LXe} enriched to 90\%, and the remaining depleted volume with \isotope[136]{Xe} concentration of ~3\%.

\subsection{Detector Design}\label{sec:geometry}

The design proposal of HERETIX aims to keep divergences from the proposed XLZD nominal case to a minimum. Thus, we consider an active \ac{LXe} target instrumented as a dual-phase \ac{TPC} with a drift length of 3\,m and a diameter of 3\,m. In the XLZD case this allows for a 60\,tonne target. The \ac{LXe} is surrounded by \ac{PTFE} panels which are a few mm thick. Beyond the \ac{PTFE} lie copper ring segments which provide the uniform drift field required to extract charge signals from deep within the detector volume. Signals within the active volume are detected by two arrays of \acp{PMT}, each with 1184 R-11410~\cite{XENONnT:Antochi2021:PMTtests} \acp{PMT}, used in XENONnT and LZ, where the \acp{PMT} are hexagonally packed. A thin \ac{GXe} layer is included between the \ac{LXe} target and the upper \ac{PMT} array to extract the delayed charge signals. All of this is enclosed within two stainless steel cryostat vessels with massive flanges and several support rings along the length of each vessel for structural rigidity.

Our proposal diverges from the XLZD nominal design by including a hermetically sealed inner volume containing 5\,tonnes of LXe enriched to 90\% in \isotope[136]{Xe}. This volume is also 3\,m in height, but has a diameter of 85\,cm. The remaining active volume outside this hermetically sealed inner volume contains around 56\,tonnes (see Table \ref{tab:geometry_dimensions} for exact values) of depleted \ac{LXe} at 3\% \isotope[136]{Xe}. We consider the inner container to be manufactured from 3\,mm thick sapphire. Dual-phase \ac{LXe} \acp{TPC} maintain a drift field within the \ac{LXe} volume and an extraction field in the \ac{GXe} region by placing a cathode at the bottom of the \ac{TPC}, a gate at the gas-liquid interface, and an anode a few millimetres above the liquid in the gaseous region. These are usually constructed from thin parallel wires or meshes made of stainless-steel. In our design, we maintain the assumption that the gate consists of a thin mesh, but replace the anode and cathode with parallel planes of sapphire that extend across the entire 3\,m diameter plane of the active region. The sapphire planes are assumed to be coated with conductive material in a grid pattern. 

\begin{table}[tbp]
\caption{The HERETIX proposal relies on the nominal XLZD~\cite{XLZD:2024nsu} proposal with minimal design changes to accommodate the second hermetic sapphire volume. We list the realisation of the major detector components studied in this work for clarity on our assumptions on the physical parameters of the HERETIX detector.}
\centering
\begin{tabular}{lr}
\toprule
\textbf{TPC}\\
Active LXe & 61.2\,tonnes \\
\acp{PMT}                            & $2\times1184$ \\
Drift length & 300\,cm \\
Diameter & 300\,cm \\
\midrule
\textbf{Sapphire Vessel}\\
\hspace{5mm}Enriched LXe & 4.8\,tonnes \\
\hspace{5mm}Radius & 42.6\,cm \\
\hspace{5mm}Thickness                  & 0.3\,cm \\
\midrule
\textbf{Cryostat}\\
Inner Vessel\\
\hspace{5mm}Radius             & 155\,cm \\
\hspace{5mm}Height               & 390\,cm \\
Outer Vessel\\
\hspace{5mm}Radius             & 170\,cm \\
\hspace{5mm}Height               & 400\,cm \\
Thickness                  & 1 cm \\
\bottomrule
\end{tabular}
\label{tab:geometry_dimensions}
\end{table}

An illustration of the proposed HERETIX detector geometry is shown in Figure~\ref{fig:G4_vis}, which visualizes a simplified geometry implemented in GEANT4 v11.4~\cite{GEANT4:2002zbu}. For this work we focused only on detector components known to contribute the majority of material background signals in previous generations of \ac{LXe} \acp{TPC}. A summary of the simulation's geometry parameters is shown in Table~\ref{tab:geometry_dimensions} for ease of reference.

\subsection{Material Backgrounds}

In \ac{0nu2b} searches, the most significant material background originates from $\gamma$-rays emitted in the decay of \isotope[214]{Bi} (a decay product of \isotope[238]{U}). This 2448\,keV $\gamma$-ray produces a signal that significantly overlaps with the \ac{0nu2b}-decay Q-value of 2458\,keV. A smaller contribution originates from partial energy reconstruction of the 2615\,keV $\gamma$-ray from the decay of \isotope[208]{Tl} (a decay product of \isotope[232]{Th}). 

For \ac{DM} searches using \ac{LXe} targets, material backgrounds are subdominant to other \ac{ER}-backgrounds due to the highly effective ability of xenon to shield from low-energy $\gamma$-rays. Therefore, the inner region of the \ac{LXe} target in the XLZD nominal design case is expected to have an overall material background well below the expected rate of irreducible solar neutrino-elec\-tron interactions. However, HERETIX would add additional material inside this ultra low-background volume to form the inner hermetically sealed volume. We thus carefully assess the overall impact of \ac{ER} backgrounds in the low-energy \ac{RoI} for \acp{WIMP} from \isotope[60]{Co}, \isotope[137]{Cs} and \isotope[40]{K} in addition to the those isotopes of concern in the high-energy \ac{RoI} for \ac{0nu2b}-decays.

\begin{figure*}[htbp]
    \centering
    \includegraphics[width=0.95\textwidth]{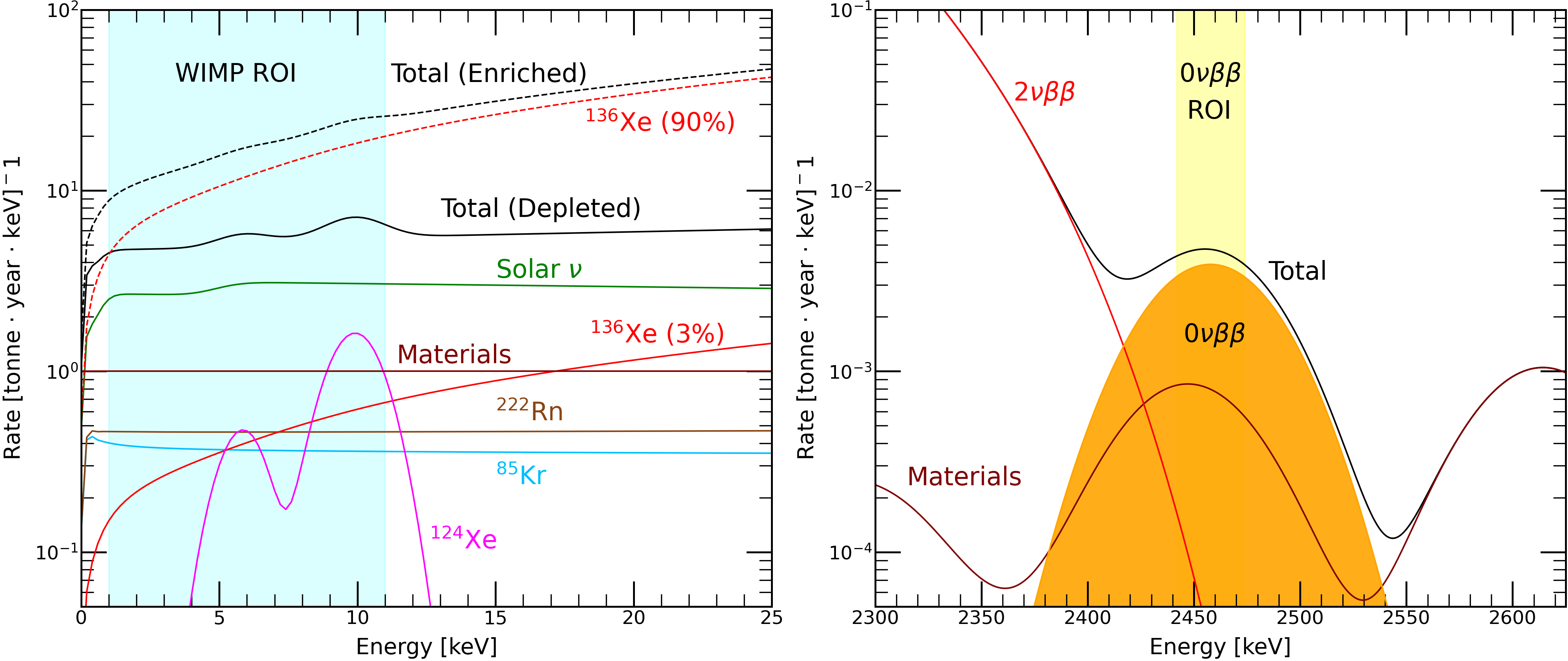}
        \caption{Energy spectra of \ac{ER} backgrounds around the WIMP and \ac{0nu2b} signal regions. Left: The low energy contributions to the \ac{ER} backgrounds for the WIMP search in the outer, depleted region. The dotted lines show the effect of a 90\% \isotope[136]{Xe} enrichment, demonstrating the high backgrounds from \ac{2nu2b} that make the inner region unsuitable for WIMP searches. Right: High \ac{ER} backgrounds in the \ac{0nu2b} Q-value region inside the enriched inner volume, with an overlaid signal corresponding to a decay half-life of $1 \times 10^{28}$ years.}
    \label{fig:ER_rates}
\end{figure*}

The assumed nominal activities of the HERETIX detector are listed in Table~\ref{tab:activities} for the most important detector components included in our GEANT4 realisation. Activities for stainless steel, PTFE and copper are taken from~\cite{XENON:2020kmp}. The activities of the R-11410 \acp{PMT} are averaged from the results reported in~\cite{XENON:2021mrg}. The activities listed in Table~\ref{tab:activities} for HERETIX adopt the same assumption as the XLZD projection, that a 75\% reduction in the \isotope[214]{Bi} rate can be achieved. This level of reduction is recently demonstrated by a new \ac{PMT} design for PandaX-xT~\cite{Wang2026}. Along with the \ac{PMT}s, the inner sapphire vessel represents a dominant contribution to the final event rate in our region of interest for \ac{0nu2b} searches. The assumed radioactivity of $^{238}$U and $^{232}$Th are reported in~\cite{Chernyak:2025fxq}. The remaining isotopes are taken from the measured activities of the ceramic (Al$_{2}$O$_{3}$) stems of R-11410 PMTs reported in~\cite{XENON:2021mrg}. We also list the total mass of each of the components in the simulations. 

\begin{table}[htbp]
\caption{
\label{tab:activities}Assumed radioactivity levels of the major detector components in HERETIX. Screening results from XENONnT are used for most components. The \isotope[238]{U} result for the \acp{PMT} in XENONnT has been reduced by 75\% in-line with the nominal case considered in the XLZD experiment. \isotope[238]{U} and \isotope[232]{Th} results for sapphire are taken from~\cite{Chernyak:2025fxq}. (1) Cryostat material: stainless steel. (2) \acp{PMT} are modelled as 3" R-11410s.}
\centering
\setlength{\tabcolsep}{3pt}
\begin{tabular}{l r l c c c c c}
\toprule
Component & Amount & Units &
\multicolumn{5}{c}{Activity [mBq/unit]} \\
\cmidrule(lr){4-8}
&
&
&
$^{238}$U &
$^{232}$Th &
$^{60}$Co &
$^{137}$Cs &
$^{40}$K \\
\midrule
Cryostat$^{1}$ & 12900  & kg   & 2.49 & 0.26 & 7.08 & 0.84 & 3.48 \\
Sapphire       & 250    & kg   & 0.03 & 0.01 & 0.99 & 0.11 & 0.21 \\
PTFE           & 290    & kg   & 0.12 & 0.11 & 0.05 & 0.04 & 2.40 \\
Copper         & 500    & kg   & 0.69 & 0.03 & 0.11 & 0.02 & 0.29 \\
PMT$^{2}$      & 2368   & piece & 2.00 & 0.45 & 0.99 & 0.12 & 14.1 \\
\bottomrule
\end{tabular}
\end{table}

\subsection{Electronic Recoil Backgrounds}\label{sec:er_bkg}

The expected \ac{ER}-background originates both from the detector materials and radio-isotopes intrinsic to the \ac{LXe} target, primarily from other noble gases and their decay products. We estimate the total rate of \ac{ER}-\-back\-grounds from materials by following the procedure described in~\cite{XENON:2015gkh} where we simulate decays, in GEANT4, of each isotope listed in Table~\ref{tab:activities}. Decays are simulated from each of the major components, and expectations are corrected for the total mass and assumed activity.   

\begin{figure*}[t!]
	\centering
    \includegraphics[width=0.95\textwidth]{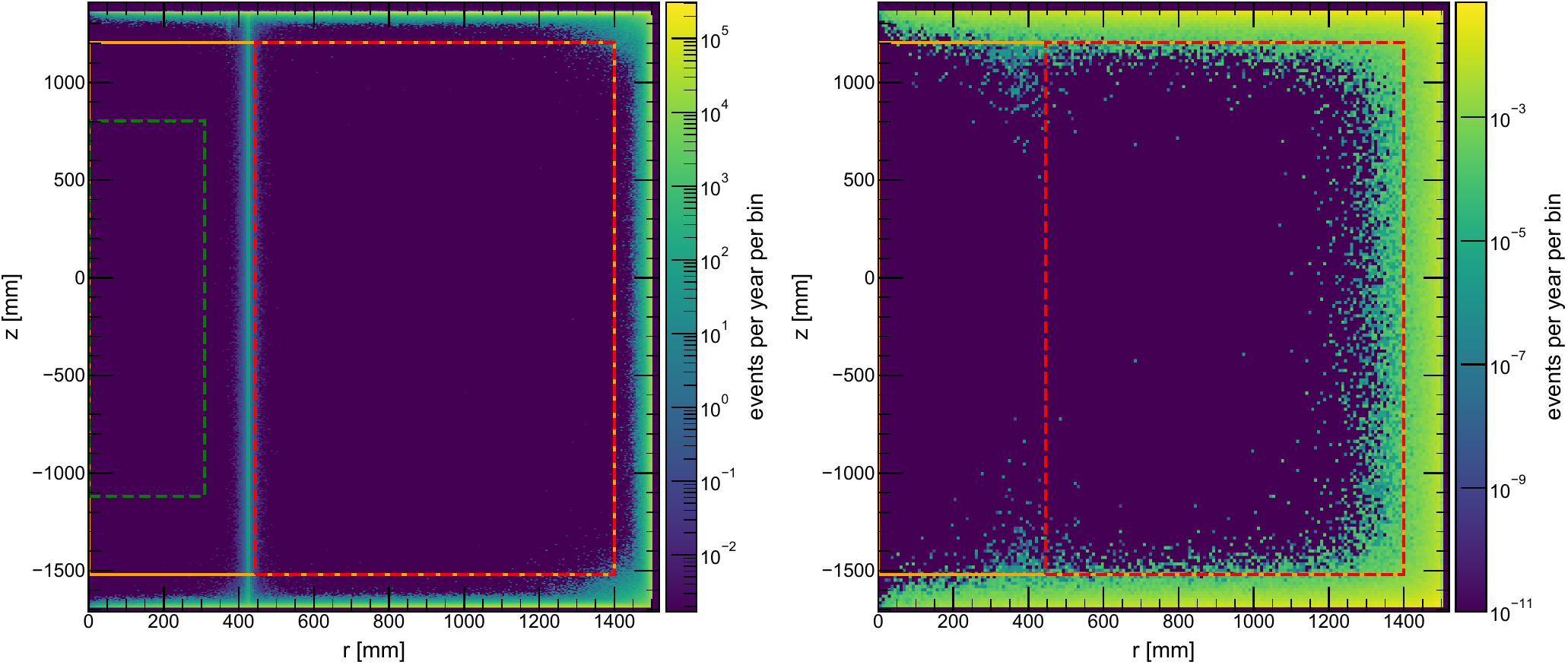}
	\caption{Spatial distribution of the low-energy \ac{ER} (left) and \ac{NR} (right) material background contributions from single scatter interactions for the \ac{WIMP} search. The outer dimension of the \ac{WIMP} search \ac{FV} with a mass of 43.2\,tonnes (red dashed) is optimised by the \ac{NR} rate, excluding the inner sapphire vessel volume which contains enriched xenon and leads to a high \ac{ER} background rate. The increased \ac{ER} material background rate at the sapphire wall is also excluded. The nominal XLZD \ac{FV} of 48.0\,tonnes (orange) and the \ac{0nu2b} search \ac{FV} (green dashed) is given for reference.}
    \label{fig:material_rate_sapphire}
\end{figure*}

As \ac{WIMP}s are expected to have very small \ac{WIMP}-nucleon cross-sections, and \ac{0nu2b} decays produce two electrons with sub-mm mean free paths, candidate events in both \ac{RoI}s are expected to topologically interact at only one location within the detector volume. High energy $\gamma$-rays from detector materials or intrinsic sources within the \ac{LXe} target can Compton scatter multiple times and depending on detector response be reconstructed as "multiple scatters" or "single scatters." Thus for background suppression from materials the spatial resolution plays an important role in the projected sensitivity. We apply a clustering algorithm to the simulated energy depositions and, in line with the XLZD nominal case, assume a uniform ability throughout the detector to resolve energy depositions separated by more than 3,mm in the z-direction. The impact of separation in the horizontal plane is assumed to be negligible. The resulting distribution of single scatter \ac{ER} events from material backgrounds is shown in the left panel of Figure~\ref{fig:material_rate_sapphire}, which illustrates the impact of the sapphire in the inner volume of the \ac{LXe} target. We select an optimised \ac{FV} such that the total  single scatter \ac{ER}-rate in the \ac{WIMP} \ac{RoI} (1.5-13.5\,keV$_\mathrm{er}$) is <1 event/tonne.year.keV. The \ac{WIMP} \ac{RoI} is chosen to match the 4-50\,keV$_\mathrm{nr}$ region which is expected to contain 72\% of the expected recoil spectrum of a 40\,GeV/c$^{2}$ mass \ac{WIMP}. This threshold is chosen such that the contribution from materials is subdominant to the irreducible neutrino-electron interaction rate. Two \ac{FV}s,  inside the hermetic cylinder and in the outer volume, are selected to exclude the region near the sapphire, and are represented by the green and red dashed lines in Figure~\ref{fig:material_rate_sapphire} respectively.

Intrinsic \ac{ER} backgrounds that are relevant to the \ac{WIMP} search originate primarily from the $\upbeta$-decay to the ground state of \isotope[214]{Pb} (a decay product of \isotope[222]{Rn}), the $\upbeta$-decay of \isotope[85]{Kr}, the \ac{2nu2b} decay of \isotope[136]{Xe} and the \ac{DEC} in \isotope[124]{Xe}. For the two xenon isotopes considered, we assume natural abundance in the case of \isotope[124]{Xe} and include captures from all electron shells that produce more than 1\,event per tonne-year. The \isotope[136]{Xe} abundance is distributed such that the inner (outer) region has 90\% (3\%) abundance. Contributions from the other noble gas isotopes are assumed to match those of the XLZD nominal case, i.e. 0.1\,$\upmu$Bq/kg of \isotope[222]{Rn} and 0.03\,ppt \isotope[\mathrm{nat}]{Kr}. In the left panel of Figure~\ref{fig:ER_rates} the total \ac{ER} background in the low energy \ac{RoI} is shown along with the individual contributions listed here. The solid (dashed) lines represent the background rate in the inner (outer) volume due to the differing \isotope[136]{Xe} concentrations. The total low-energy \ac{ER} rate in the inner volume in the case of a 90\% enriched target would result in a total \ac{ER} rate in the \ac{RoI} on-par or in excess of that achieved in the current generation of \ac{LXe} based \ac{DM} detectors. Thus we exclude this volume for the \ac{WIMP} search in our nominal proposal. 

The \ac{0nu2b} \ac{RoI} is defined as a 1$\sigma$ region around the Q-value (\qvalue). We assume an energy resolution on-par with the XLZD nominal case of $\sigma_E$\,=\, 0.65\%~\qvalue\,and construct a 32\,keV \ac{RoI} defined by \qvalue\,$\pm\,1\sigma$. The \ac{FV} is primarily constrained by \isotope[214]{Bi} decays, the most significant of which originate in the radial dimension from the sapphire and in the vertical direction from the \ac{PMT}s. The \ac{FV} inside the inner sapphire cylinder is optimised using a figure-of-merit estimator using a counting experiment model, and results in a \ac{FV} for the \ac{0nu2b} search of 2.80\,tonnes. This background and the expected signal from a \ac{0nu2b}-decay with a half-life of $1 \times 10^{28}$\,years are shown in the right plot of Figure~\ref{fig:ER_rates}, along with the expected signal from \ac{2nu2b} decays in a 90\% enriched \ac{LXe} target. 

\subsection{Nuclear Recoil Backgrounds}\label{sec:nr_bkg}

Low-energy elastic scattering of radiogenic neutrons, muon-induced neutrons, and neutrons produced by cosmogenic and solar neutrino interactions, along with \ac{CEVNS} contribute to the \ac{NR} background in our \ac{WIMP} low-energy \ac{RoI} (4-50\,keV$_\mathrm{nr}$). In the \ac{0nu2b} \ac{RoI} contributions from \ac{NR}s are nonexistent. We neglect other detector specific components such as accidental coincidences or gamma-X~\cite{XENON:2020kmp} events in this work.

Among the low energy \ac{NR} backgrounds, muon-induced neutrons and radiogenic neutrons from the laboratory are expected to be reduced to negligible levels by the experiment's water shielding, in line with the XLZD proposal. Furthermore the XLZD experiment aims to suppress the background from radiogenic neutrons to below the irreducible neutrino backgrounds of about one event over the full exposure of the proposed experiment~\cite{XLZD:2024nsu}.

\begin{table}[t]
\centering
\captionsetup{skip=8pt}
\caption{Optical material parameters used in the GEANT4 simulation taken from~\cite{Althueser2023,XENON:2020kmp,RefractiveIndexSapphiremeller,RefractiveIndexSapphireTydex}. The estimate of the sapphire absorption length is discussed in Section~\ref{sec:innermaterial}.}
\label{tab:geantoptics}
\begin{tabular}{lr}
\toprule
Parameter & Value\\
\hline
\ac{LXe} refractive index & $1.63$ \\
\ac{LXe} absorption length & $50$\,m \\
\ac{LXe} scattering length & $50$\,cm \\
\midrule
\ac{GXe} absorption length & $100$\,m \\
\ac{GXe} scattering length & $100$\,m \\
\midrule
Sapphire refractive index & $1.92$\\
Sapphire absorption length & $28.5$\,mm - $30$\,m\\
\midrule
\ac{PTFE} reflectivity & $99$\,\% \\
Electrode transmittance & $96$\,\%\\
SS reflectivity & $20$\,\%\\
\bottomrule
\end{tabular}
\end{table}

The radiogenic neutrons produced through spontaneous fission or ($\upalpha$,\,n) reactions in the detector materials represent the major component of single scatter \ac{NR} interactions within the \ac{LXe}. We estimate the neutron yields and energies originating from the detector components using the SOURCES-4A~\cite{madland1999sources4a} software as described in~\cite{XENON:2015gkh} by the XENON collaboration. The emission of a single neutron is conservatively simulated, neglecting coincident gamma rays. Here, the $\upalpha$-particle originates from \isotope[238]{U}  and \isotope[232]{Th}-chain, and the neutron interaction in the \ac{LXe} target are selected and weighted by the neutron yield and material activity as given in Table \ref{tab:activities}. As with the \ac{ER} backgrounds, only single-scatter interactions within the \ac{FV} contribute to the \ac{WIMP} search background, where the same clustering is used as described in Section~\ref{sec:er_bkg} to distinguish between multiple and single scatter interactions. The final single scatter \ac{NR} background is shown in the right panel of Figure~\ref{fig:material_rate_sapphire} inside the full active volume.

The nominal XLZD sensitivity calls for a \ac{NR} background rate in our \ac{WIMP} \ac{RoI} of <0.5 events per tonne-year. In isolation from \ac{ER} background concerns, we optimise a \ac{NR} background region only, in order to understand the impact of the proposed inner \ac{TPC} volume. As is evident from Figure~\ref{fig:material_rate_sapphire}, the impact of the additional sapphire material with regards to raising the overall \ac{NR} rate is negligible. This can be attributed to the likelihood of a neutron emitted from the sapphire scattering only once in the approximately 1\,m of xenon between the sapphire and the outer edge of the active volume being highly suppressed. Therefore, we can construct a \ac{FV} with a total mass of 48.0\,tonnes for the \ac{WIMP} search which is agnostic to the presence of the sapphire fulfilling the rate requirement for the radiogenic neutrons. However, for the \ac{WIMP} search the entire inner volume of the sapphire vessel is excluded due to other backgrounds as outlined in Section~\ref{sec:physics}.

\begin{figure}[t]
	\centering
	\includegraphics[width=0.5\textwidth]{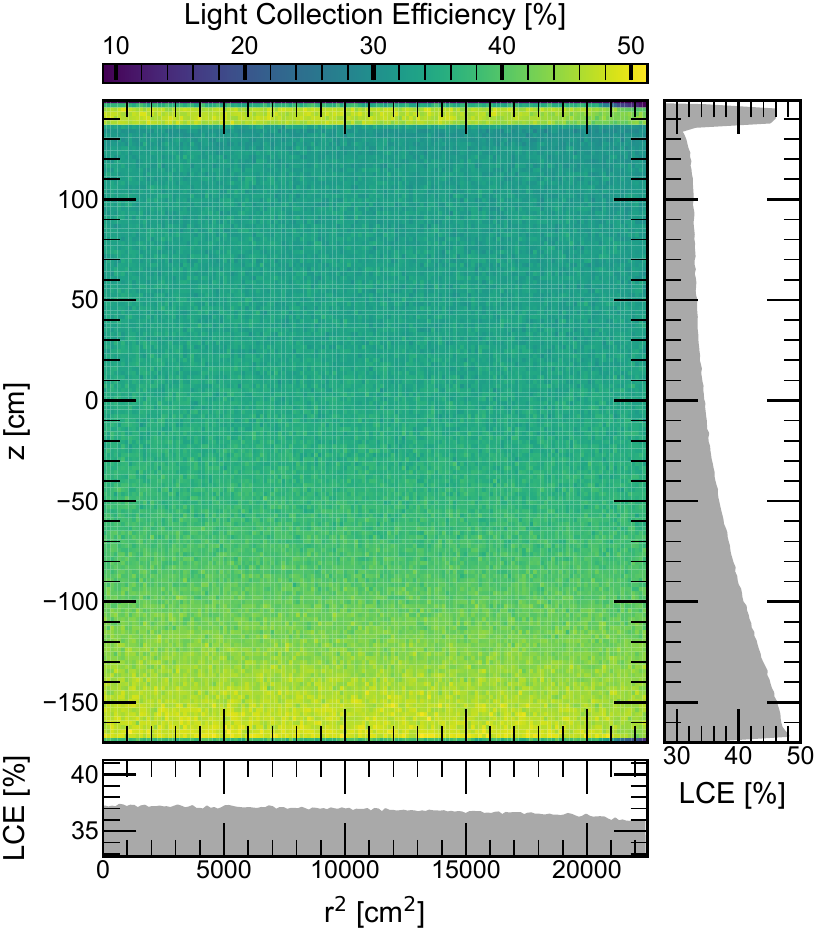}
	\caption{Light collection efficiency (\acs{LCE}) of HERETIX estimated from optical simulations with GEANT4. The \acs{LCE} is highest in the gas gap above the liquid level at the top of the \ac{TPC} and lowest directly below the liquid-gas interface. It increases with decreasing distance to the bottom \acp{PMT} in the \ac{LXe}. The sapphire vessel does not lead to localised features as discussed in Figure~\ref{fig:efficiency_wimp}.}
    	\label{fig:LCEmap}
\end{figure}

\subsection{Detector Response}\label{sec:detector_response}

In dual-phase \ac{LXe} \ac{TPC}s, \ac{0nu2b} searches typically rely on the combined energy scale (CES) which combines the primary (S1) and secondary (S2) light signals to reconstruct the deposited energy of an interaction~\cite{PhysRevB.68.054201}. Therefore, in this work we assume that the simulated deposited energy in a single scatter interaction corresponds to the measured energy scale in our nominal proposal. To account for broadening of mono-energetic signals due to \ac{LCE}, recombination fluctuations in the number of photon and electrons produced at the interaction vertex, and electron loss due to drift time and extraction efficiency, we apply an energy smearing as described by XENON1T ~\cite{Aprile2020EnergyResolution}.

\ac{WIMP} searches, relying on \ac{NR} signals, typically occur in the two-dimensional S1-S2 parameter space. In the case of S1 signals, the low-energy nature of \ac{WIMP} searches occurring right at the detection threshold, places a large importance on the overall \ac{LCE} within the active detector volume. Therefore, we devote effort to studying the \ac{LCE} in the HERETIX detector, and propagate this through to our expected S1-S2 response model for the major background components and various \ac{WIMP} masses.

We simulate the light collection in our nominal proposal with the central sapphire structure, but also consider the XLZD nominal design without the hermetic \ac{TPC} volume in order to have a like-to-like benchmark for comparisons. The relevant material properties for optical simulations are shown in Table~\ref{tab:geantoptics}. The refractive index of sapphire has been measured at a wavelength of $200$\,nm by multiple companies for optical components~\cite{RefractiveIndexSapphiremeller,RefractiveIndexSapphireTydex} and was found to be $1.92$. However, the absorption length of sapphire for VUV light in a LXe environment has not been measured. We assume an absorption length similar to the quartz material from the \ac{TPC} \acp{PMT} of $30$\,m, corresponding to an absorption of $0.01$\,\%. Alternative estimations indicate absorption lengths of down to 28.5\,mm and will be discussed in Section~\ref{sec:innermaterial}. The \ac{LCE} map for HERETIX is shown in Figure~\ref{fig:LCEmap} and results in average \ac{LCE} of 36.6\,\% in the active region and 37.2\,\% (36.5\,\%) in the inner (outer) volume.

In order to simulate our expected S1 signals in the low energy region, we rely on photon detection efficiency g$_1$, which combines the \ac{LCE} effects with the quantum efficiency and collection efficiency at the first dynode of the \ac{PMT}s. The g$_1$ value is calculated using the simulated \ac{LCE} map following the procedure outlined in~\cite{Althueser2023}. The quantum efficiency of the Hamamatsu R11410-21 \acp{PMT} has been measured to be $0.34\pm 0.03$ at room temperature on average~\cite{QuantumEfficiency} and increases by $10$\,\% at cryogenic temperatures~\cite{QuantumEfficiency}. Here, the double photoelectron emission of $22\%$ is already included in the quantum efficiency~\cite{QuantumEfficiency}. The collection efficiency depends on the position of the photon on the photocathode and is dependent on the incident angle~\cite{XENONnT:Antochi2021:PMTtests}. A uniform distribution and isotropic angular distribution is assumed, resulting in a collection efficiency of $1$. The average g$_1$ value in the central detector volume is found to be 0.146 PE/ph in the nominal case.

For the detector response simulation, it is assumed that the detector is filled with \ac{LXe} at 2\,bar pressure with a temperature of 177\,K. The signal response for energy depositions in \ac{LXe} is simulated using the \ac{NEST} parametrisation, describing the excitation and ionisation processes as functions of both energy and electric field. It gives an estimate on the number of photons and electrons for a certain energy deposit and recoil type. The photons of the S1 signal are distributed on the \acp{PMT} with respect to the photon detection efficiency g$_1$ and a three-fold \ac{PMT} coincidence requirement. A single photoelectron acceptance of 93\%~\cite{XENON:2024wpa} and a resolution of 25.1\%~\cite{XENONnT:Antochi2021:PMTtests} is assumed for the \acp{PMT}. Afterwards, a \ac{PMT} channel specific trigger threshold of 0.13\,\ac{PE}~\cite{XENON:2023cxc} is considered. The S1 parameter space is restricted to (3, 100)\,\ac{PE}. Electrons drift within the \ac{LXe} volume, with losses accounted for by an electron lifetime of 10\,ms. The drift field is assumed to be homogeneous, neglecting electric field defects at the detector walls, and is set to 240\,V/cm providing an optimal discrimination between \ac{ER} and \ac{NR} recoils~\cite{LUX:2020car}. The electrons are then extracted into the gas region by an extraction field of 6\,kV/cm~\cite{XLZD:2024nsu}. S2 photons are distributed on the \acp{PMT} with an average detection efficiency g$_1$ in the gas gap of 0.16 PE/ph. The S2 parameter space ranges from (1, 8000)\,\ac{PE}.

The efficiency of reconstructing an \ac{ER} or \ac{NR} event considering the detector response model with a certain total energy is shown in Figure~\ref{fig:efficiency_wimp}. No selection acceptance is considered. The \ac{WIMP} search \ac{RoI} with at least 50\% efficiency is therefore defined as (1.5, 13.5)\,keV$_\mathrm{ER}$ and (4, 50)\,keV$_\mathrm{NR}$, leading to an average efficiency of 93.3\% and 93.4\%, respectively.

\section{Projected Sensitivity}\label{sec:physics}

The estimated physics reach for the two rare event searches is determined using the profile likelihood approach. We convert the backgrounds considered in Section~\ref{sec:backgrounds} into 2-dimensional S1-S2 templates for the \ac{WIMP} search and 1-dimensional \ac{CES} templates for the \ac{0nu2b} search. The \ac{0nu2b}-search is performed in a fiducial volume of 2.8\,tonnes and we consider exposures of up to 12 years. 

\begin{figure*}[t]
	\centering
	\includegraphics[width=0.95\textwidth]{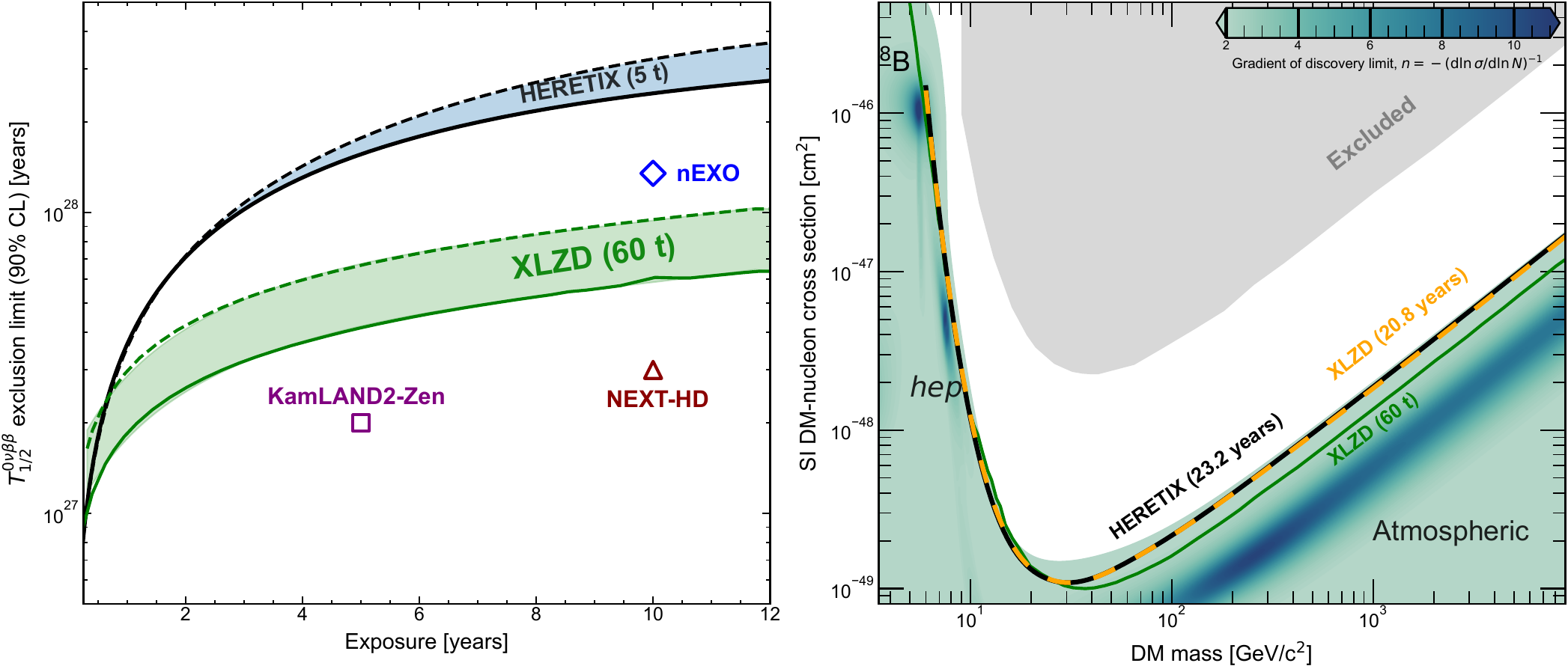}
	\caption{Left: Projected exclusion limits of HERETIX for the half-life of \ac{0nu2b} of \isotope[136]{Xe}. Shown in solid lines, both for our proposal and XLZD, are the nominal cases with the background and detector response assumptions detailed in Section~\ref{sec:backgrounds}. The dashed lines represent optimistic assumptions on detector backgrounds. We also show the projections of dedicated future \ac{0nu2b}-searches: nEXO~\cite{nEXO:2021ujk}, KamLAND2-Zen~\cite{Grant2020KamLANDZEN} and  NEXT-HD~\cite{NEXT:2020amj}. Right: Projected 90\%\,C.L. upper limits on the spin-independent WIMP-nucleon cross section for 1000\,t.y exposure for our XLZD-like detector (orange) and excluding the HERETIX volume (black). The XLZD exclusion limit~\cite{XLZD:2024nsu} (green) is given as reference. The blue shaded regions indicate the neutrino fog as defined in\,\cite{Carew:2023qrj} following the conventions from\,\cite{Baxter2021}. The gray shaded region shows the already excluded parameter space by current experiments~\cite{lzwimp2025,xenonntwimp2025}.}
    	\label{fig:combined_limits}
\end{figure*}

Our target exposure for the \ac{WIMP} search is set to be equivalent to the XLZD design goal of 1000\,t.y, which necessitates a 23.2 year lifetime; 2.4 years longer than for XLZD. The slightly longer lifetime is a result of the smaller volume in HERETIX as compared to the XLZD nominal case. As mentioned in Section~\ref{sec:er_bkg}, the presence of a 90\% enriched \ac{LXe} target within the sapphire volume raises the overall \ac{ER} background above 10\,events/tonne/year.keV, which is a level equivalent to the current generation of \ac{LXe} \ac{DM} detectors. Therefore, we exclude the entire inner region from the \ac{WIMP} search resulting in a \ac{FV} of 43.2\,tonnes, 10\% smaller than the 48\,tonnes assumed in the nominal 20.8 years lifetime of XLZD.

\subsection{\ac{WIMP} Sensitivity}\label{sec:wimp_sensi}

In our proposal we consider a simplified background model for the \ac{WIMP} search with four components. The \ac{ER} material backgrounds discussed in Section~\ref{sec:er_bkg} are combined into a single template in our \ac{WIMP} \ac{RoI}. The \ac{NR} backgrounds from Section~\ref{sec:nr_bkg} are treated as three different templates: one template for the radiogenic neutron backgrounds from materials, a combined template for \ac{CEVNS} from atmospheric and diffuse supernova, as well as a template for solar neutrinos including \textit{hep} and \isotope[8]{B} neutrinos. The \ac{CEVNS} templates are derived from~\cite{XENON:2020kmp}. We apply our detector response model described in Section~\ref{sec:detector_response} to these templates, as well as to our \ac{WIMP} models. 

Projected 90\% exclusion \ac{WIMP} sensitivities, with the HERETIX xenon mass removed from the fiducial target, are shown in the right panel of Figure~\ref{fig:combined_limits}. For a like-to-like comparison, we derive limits for the nominal XLZD detector using the same framework at an exposure of 1000\,ty and additionally show the projections of the XLZD collaboration from~\cite{XLZD:2024nsu}. Our re-implementation yields slightly weaker limits above 30\,GeV, which we attribute to differences in the assumed detector response as outlined in Section~\ref{sec:detector_response} and to the more detailed treatment of the optical performance adopted in this work. At equivalent exposures, the introduction of the sapphire vessel with optimal optical properties produces no meaningful degradation of the \ac{WIMP} sensitivity relative to the nominal XLZD detector. Therefore, only the correspondingly longer livetime required to accumulate the same exposure with the reduced fiducial mass needs to be accounted for.

\subsection{\ac{0nu2b} Sensitivity}\label{sec:0nu2b_sensi}

Background constraints from \isotope[214]{Bi} decays in the sapphire and the \ac{PMT}s restrict the usable \ac{FV} region in the inner volume to 2.80 tonnes as shown by the green dashed line in the left panel of Figure~\ref{fig:material_rate_sapphire}. We consider two backgrounds in our projected sensitivity, the contribution of \ac{ER} backgrounds from materials and the rate of \ac{2nu2b} \isotope[136]{Xe} decays. These are fit in a broad energy region around \qvalue~to constrain their contributions in our \ac{RoI} (\qvalue\,$\pm\,1\sigma$).

Shown in Figure~\ref{fig:combined_limits} is the projected exclusion limit at 90\% confidence limit for the half-life of the \ac{0nu2b} decay of \isotope[136]{Xe} as a function of livetime. Here we neglect uncertainties in the absorption of UV light in sapphire as the energy signals are sufficiently large to be unbiased by any \ac{LCE} effects. Also shown are the projected exclusion limits from the nEXO~\cite{nEXO:2021ujk}, KamLAND2-Zen~\cite{Grant2020KamLANDZEN} and  NEXT-HD~\cite{NEXT:2020amj}. The nominal case for both our proposal and XLZD, assuming a 75\% reduction in \isotope[214]{Bi} rates in the \ac{PMT}s compared to the measured value in current large \ac{LXe} detectors, is shown in solid lines for a 12\,year exposure. The dashed lines represent further optimistic scenarios where it is possible to reduce the \isotope[214]{Bi} by 90\% in the \ac{PMT}s and the measured \isotope[238]{U} activity of the sapphire to below  1\,$\mathrm{\upmu Bq/kg}$.

HERETIX will be able to leverage the larger depleted \ac{LXe} surrounding the hermetic volume to achieve much lower backgrounds in the \qvalue\, \ac{RoI} compared to the projected background of the nEXO experiment. This increases the sensitivity even in the case of the simple 1-D likelihood model considered here. After a 4.1 (3.7) year exposure the nominal (optimistic) proposal will be able to exclude, at 90\% \ac{CL}, half-lives of up to 1.35$\times10^{28}$ years, exceeding the projected nEXO sensitivity. After a 10\,year exposure the 90\% \ac{CL} is projected to reach 2.5$\times10^{28}$ years (3.2$\times10^{28}$ years). This is an improvement of 4.1 times on the XLZD proposal, where the combination of the enhanced signal from the enriched xenon target and the very clean background in our hermetic \ac{FV} provides much greater sensitivity after only 1\,year of exposure. In the optimal HERETIX scenario we project sensitivities 2.5 times of the nEXO projection, due to the larger \ac{FV} enabled through the improved shielding of the hermetic volume.

\section{Technical Implementation}\label{sec:detector}

HERETIX, as described in Section~\ref{sec:backgrounds}, provides a clear scientific case that allows two rare event searches to be pursued within a single framework. As shown in Section~\ref{sec:physics}, HERETIX will reach world leading sensitivities for the half-life of the \ac{0nu2b}-decay of \isotope[136]{Xe}, while incurring only a modest penalty on the required exposure to achieve the planned \ac{WIMP} sensitivity of the nominal XLZD proposal. 

The concept of combining multiple targets in a single xenon detector was previously explored in the multi-target XAX proposal~\cite{Arisaka2008}, utilising an \isotope[136]{Xe}-enriched inner detector volume forming a hermetically sealed inner detector. This principle of a hermetic inner detector was further explored in the GraXe design using a graphene balloon~\cite{GomezCadenas2011}, outlined for LBECA in~\cite{Bernstein2020} for a fused silica vessel, and experimentally investigated in~\cite{Sato:2019qpr} using quartz and, more recently, in~\cite{Dierle2023} utilising \ac{PTFE}, motivated by the need to suppress radiogenic backgrounds by isolating the target from surfaces that emanate \isotope[222]{Rn}, for example. The first experimental work employed a small dual-phase xenon chamber with a quartz vessel (48\,mm diameter, 58\,mm drift length), chemically etched stainless-steel grids as anode and cathode, a gold-plated wire gate, and field-shaping rings. This prototype successfully demonstrated that small-scale hermetic containment can be achieved without degrading detector performance. The second followed the well-established design of leading xenon \acp{TPC}, utilising an inner \ac{PTFE} cylinder, relying on volume separation by cryofitting, and introducing an active \ac{LXe} level-control system.

However, despite offering clear advantages for background mitigation, several technical challenges must be addressed in order to bring the proposed detector to fruition and allow for functionality as a \ac{LXe} \ac{TPC}. The enriched inner target is naturally shielded by the depleted xenon and by isolating the inner target volume from direct contact with external materials it is possible to significantly reduce radon-induced backgrounds in the liquid volume and minimise long-lived plate-out contamination. 

\subsection{Instrumentation as dual-phase \ac{TPC}}

\begin{figure}[tbp]
	\centering
	\includegraphics[width=0.4\textwidth]{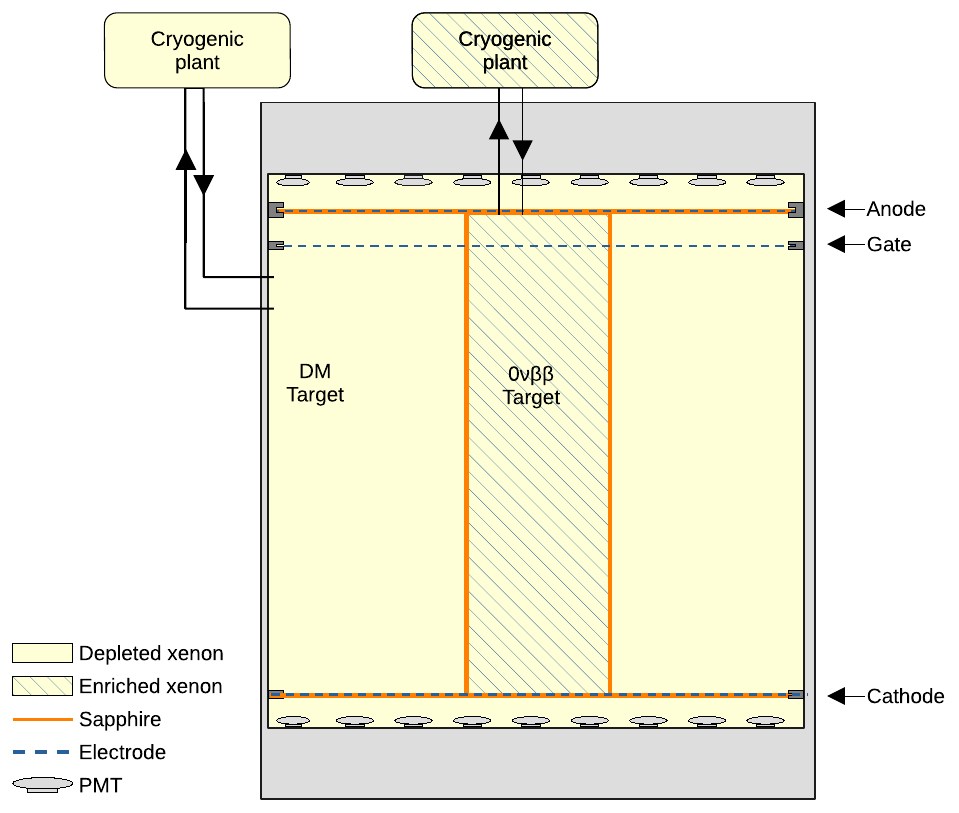}
	\caption{Conceptual layout of the HERETIX proposal. The XLZD cryostat is shown in grey, with the light orange TPC volume inside. The depleted xenon region serves as the \ac{DM} target, while the inner sapphire-contained volume holds the \isotope[136]{Xe}-enriched \ac{0nu2b} target. Two independent xenon-handling plants are shown schematically: one for depleted xenon and one for enriched xenon, each providing cryogenics, purification, calibration, distillation, recovery, and storage. The electrode implementation is conceptual: electrodes may be fully integrated on the sapphire surface, as illustrated here for the anode and cathode, or only partially integrated on the sapphire surface, e.g., only covering the inner sapphire cylinder. Mechanical support of the inner vessel remains to be defined, and may be provided by support feet from below, not shown, and/or by support through the sapphire electrodes.} 
    	\label{fig:conceptual}
\end{figure}

In HERETIX both volumes act as a dual phase \ac{TPC}. This requires the establishment of a drift and extraction field across both volumes. These fields are traditionally maintained by stainless steel wire or mesh grid electrodes which extend across the entire x-y plane of the detector. As \ac{LXe} \acp{TPC} have increased in size the manufacturing of sufficiently large grids has become more challenging, especially in the case of the gate and anode grids which establish the extraction field. Typically these grids are separated by $\mathcal{O}$(10\,mm), and have strict planarity requirements in order to maintain uniform extraction probability and single electron response across the detector. In XLZD scale detectors, the ability to maintain sufficiently planar electrodes across a 3\,m diameter remains unproven. Larger electrodes may require more robust wire diameters or mesh elements, thus reducing the overall transparency of the electrode grid with respect to allowing the xenon scintillation light to reach the photo-sensors.

HERETIX requires two hermetically separated volumes in order to maintain the enriched nature of the central volume, which poses additional challenges with implementing wire or mesh grids in the central volume. For that reason our proposal implements solid sapphire plane electrodes for the cathode and the anode that extend across both volumes. These sapphire planes can be bonded to the central cylinder. We envision that the electrodes in HERETIX would consist of thin layers of conductive material deposited onto the sapphire. As the sapphire itself will provide the structural rigidity to maintain planarity the conductive strips can be optimised to provide maximum transparency. Implementation of the gate electrode in the central volume poses a technical challenge that requires hermetically integrating wires or a metal assembly into the central volume to avoid blocking electron extraction into the gas phase; however, the total span of wires in the gate electrode in both volumes will be smaller than the current generation of \ac{LXe} detectors, and thus technically feasible from a planarity viewpoint. If necessitated, conservative approaches using regular anode and gate meshes outside of the inner container could also be realised. This approach would still reduce the required structural integrity of the wire meshes by using the inner container to span across the horizontal plane of the outer detector.

\begin{figure*}[t]
	\centering
	\includegraphics[width=0.95\textwidth]{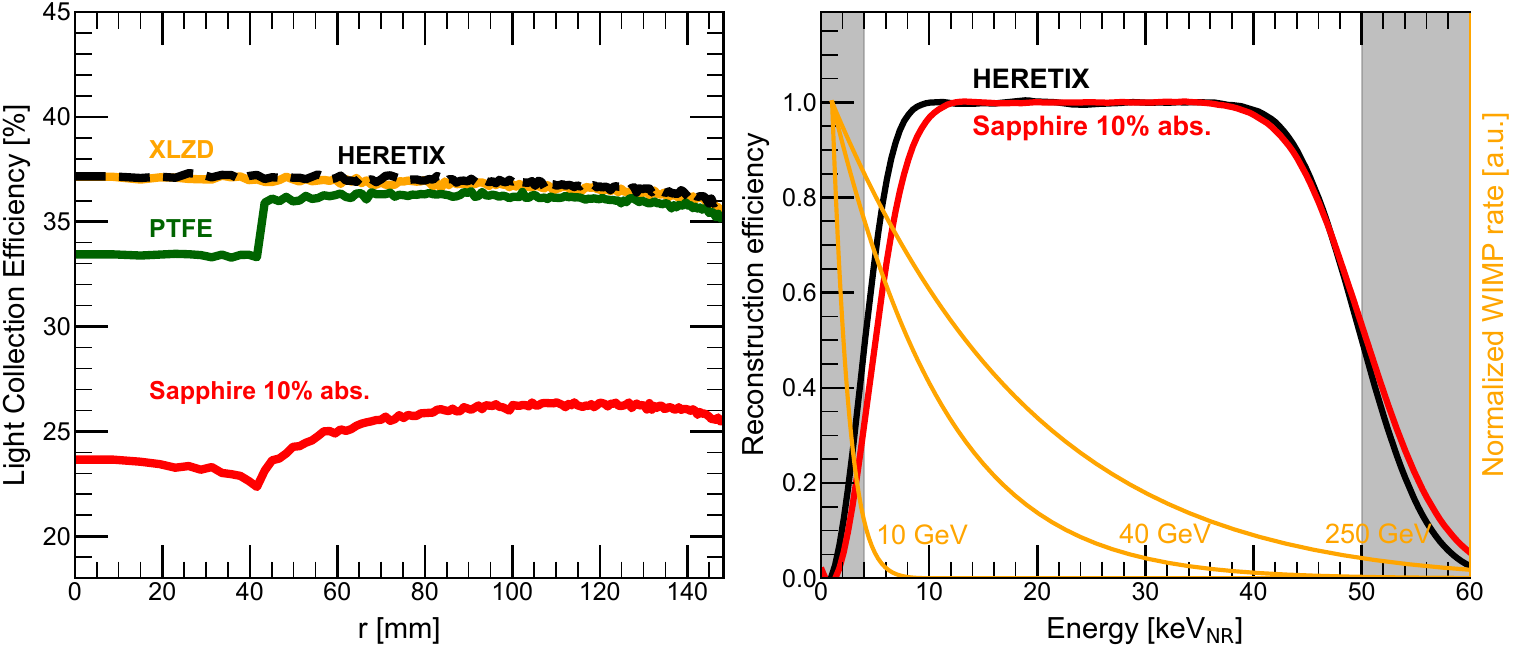}
	\caption{Left: Average \ac{LCE} along the radius of the \ac{TPC} for our implementation of XLZD (orange) and for various HERETIX vessel materials: Sapphire as discussed earlier (black), Sapphire with 10\,\% absorption (red) and for PTFE (green). Right: Reconstruction efficiency of \ac{NR} events considering the \ac{NEST} parametrisation of the detector response model. The (4, 50)\,keV$_{\mathrm{NR}}$ region with at least 50\% efficiency considering the HERETIX volume is highlighted and identical to our XLZD implementation. This region is reduced to (5.3, 50)\,keV$_{\mathrm{NR}}$ for Sapphire with 10\,\% absorption.}
    	\label{fig:efficiency_wimp}
\end{figure*}

In addition to coating the cathode and anode planes to act as electrodes, consideration should also be given to developing suitable coatings, and associated electrical connections, on the cylindrical walls of the central sapphire vessel. This vessel provides, on both faces combined, $\mathcal{O}$(15\,m$^2$) surface area for charge accumulation which could introduce complicating backgrounds in the \ac{RoI} for \ac{WIMP} searches. Charge build up over time, if not mitigated, would also distort the uniformity of the drift field. This requires a dedicated experimental campaign to study the behaviour of electrical fields produced by such sapphire vessels and electrodes in a \ac{LXe} detector. For example, a few nanometer thick coating of TiO$_2$ deposited via atomic layer deposition would give the vessel surface a layer of very high ohmic electrical conductivity. This layer could be used to mitigate charge-up and provide a homogenic electric field for the inner vessel~\cite{Goenner2022}. Alternatively a single spirally wrapped high impedance wire could be used. Both methods would not require additional support structures and thus incur negligible additional backgrounds.

\subsection{Single Phase Central Detector}

The XLZD nominal design case implements a dual phase \ac{LXe} \ac{TPC}, as the amplification and near perfect detection of scintillation light from extracted electrons provides excellent performance near threshold and decent discrimination between \ac{ER}s and \ac{NR}s. Proposals which prioritise the \ac{0nu2b} search region have traditionally favoured optimal energy resolution around \qvalue, and thus relied on collecting the charge directly to reduce fluctuations in the number of quanta which broaden the response. Given the large background in the low energy \ac{RoI} which complicates the \ac{DM} search, an alternative scenario of a single phase central volume could be considered. In this scenario a grid electrode in the central volume is no longer required, which simplifies the technical implementation in this regard, but would require instrumented readout of a wire plane in the central sapphire volume to collect charges. Additionally the central volume can be filled without the need to maintain the liquid level in the extraction region of a dual-phase detector. The additional small \ac{LXe} volume is unlikely to provide meaningful shielding from backgrounds from the top PMT array, but the density of photo-sensors in the central region on the top array could also be reduced to an optimum level which could enlarge the usable target for the \ac{0nu2b}-search. While the reduction in photo-sensors reduces the \ac{LCE} of the detector, the impact is primarily concentrated to the inside of the inner volume. Even when removing all PMTs in the central region, the average \ac{LCE} decreases only slightly from 36.6\,\% to 32.8\,\% in the active region arising from a reduction of 7\,\% (3.5\,\%) in the inner (outer) volume.

\subsection{Inner Volume Material}\label{sec:innermaterial}

The choice of vessel material is critical for the proposed detector concept, since it directly affects optical performance, xenon containment, and intrinsic backgrounds. When considering optical performance, to minimise the impact of the proposed hermetic detector on the light collection, the vessel material must be either transparent to VUV light, such as quartz or sapphire, or highly reflective, such as \ac{PTFE}. While both options reduce the fraction of the scintillation light lost to absorption, a reflective vessel increases the number of reflections and the total path length before they reach a photo-sensor and are detected. This effect is amplified by the small diameter of the inner vessel and leads to a reduced \ac{LCE}, particularly for interactions occurring inside the hermetic detector, resulting in a strongly volume-dependent detector response, as shown in Figure~\ref{fig:efficiency_wimp}. For this reason, we restrict the material selection only to materials transparent to VUV light.

High-quality sapphire is transparent in the VUV and mechanically robust. Screening of sapphire samples~\cite{Chernyak:2025fxq,Leonard:2007uv} has shown that it can be produced with low contamination of intrinsic radiogenic backgrounds from \isotope[238]{U} and \isotope[232]{Th}, although further suppression would be needed to fully exploit the inner volume for the \ac{0nu2b} search. 

The absorption of VUV light in sapphire immersed in \ac{LXe} is not known and our nominal scenario assumes properties similar to quartz. Dedicated transmission measurements of VUV light for sapphire in LXe should be carried out to quantify its impact on the \ac{LCE}. In the absence of such measurements, estimations can be made based on transmission measurements in air reported in~\cite{RefractiveIndexSapphireTydex,AbsorptionSapphireThorlabs}. However, the high refractive index of sapphire leads to significant reflectivity at the air-sapphire interface, complicating a direct comparison. Combining the reported $70\,\%$ transmission in air for a 3\,mm thick sapphire sample at $200\,\mathrm{nm}$ with the expected $20\,\%$ reflection loss from the refractive-index mismatch, we estimate a conservative upper limit on the absorption of $10\,\%$, corresponding to an absorption length of $28.5\,\mathrm{mm}$. We estimated the photon detection efficiency g$_1$ to be reduced by 31\% to $0.1$\,PE/ph leading to a reduction of the efficiency for low energy \ac{NR} events, shown in Figure~\ref{fig:efficiency_wimp}. The 90\% exclusion sensitivity for 40 GeV/c$^2$ mass WIMPs is reduced by 11\% to $1.3\times10^{-49}$\,cm$^2$ for an exposure of 1000 ty.

Should sapphire prove incompatible with the overall \ac{LCE} requirements of HERETIX, alternative materials could be considered for the inner vessel like quartz or fused silica, as already employed in the hermetic prototype of~\cite{Sato:2019qpr}. They offer good VUV transparency, chemical compatibility with xenon, and can be fabricated with low levels of contamination. These are more challenging materials from a structural integrity viewpoint, as sapphire is three to five times more resistant to crack propagation, five times stiffer than quartz, and can sustain four to ten times as much load before bending. In addition, the thermal conductivity of sapphire is an order of magnitude higher than that of quartz, which is advantageous for stable operation at cryogenic temperatures and for the passive thermal coupling of the inner vessel to the surrounding LXe bath discussed in Section~\ref{sec:xenon_handling}.

\subsection{Enrichment strategy}
The isotopic enrichment is typically performed by a cascade-based technique such as gas centrifugation. As this represents a non-trivial endeavour the choice of the total enriched mass and targeted \isotope[136]{Xe} concentration deserves careful consideration. We assume that any enrichment of the inner volume leads to a depletion of the outer volume, conserving the total xenon content. The processing effort, defined as the product of throughput mass and number of stages, scales strongly non-linearly with the target concentration: Early enrichment of the large xenon mass is expensive while late enrichment of a reduced mass is cheap. As a benchmark reaching 30\% enrichment from natural abundance would represent two-thirds of the total effort required to reach 90\% enrichment.

\begin{figure}[t]
    \centering
    \includegraphics[width=0.45\textwidth]{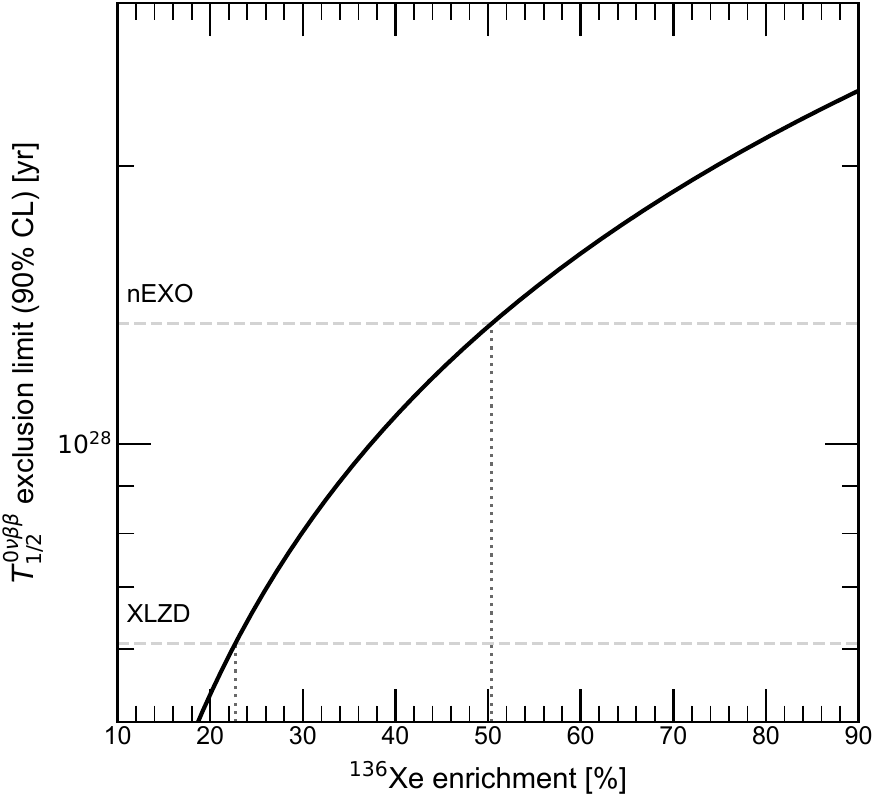}
    \caption{
    Expected 90\% C.L. exclusion sensitivity to \ac{0nu2b}-decay in our nominal proposal as a function of xenon enrichment, assuming the median excluded signal rate obtained from the detector sensitivity study after 10 years of operation. The nominal scenario considered in this work, corresponding to 90\% enrichment, is indicated by the black marker at the upper end of the curve. The horizontal dashed lines indicate the projected sensitivities of nEXO ($1.35\times10^{28}$ yr) and the XLZD 60\,tonne case ($6.09\times10^{27}$ yr). The corresponding vertical dotted lines mark the enrichment fractions at which the projected detector sensitivity reaches these benchmarks. 
    }
    \label{fig:enrichment_sensitivity}
\end{figure}

Here, we compare several strategies for achieving the targeted enrichment of the 4.8\,t inner volume in comparison to the direct enrichment to 90\% \isotope[136]{Xe}, assuming a total xenon inventory of 80\,t for XLZD. Producing 4.8\,t of 30\% \isotope[136]{Xe} inventory, by processing around 16\,t natural xenon to 90\% and diluting with the depleted by-product, costs only one-fifth of the effort of the enrichment to 90\%. A staged plan with an enrichment to 30\% and re-processing to 90\% would cost one-third more than going directly to 90\% since part of the early costly enrichment effort is effectively spent twice when the diluted intermediate by-product is re-concentrated. Despite the increased effort, a staged approach can nevertheless be scientifically attractive as the initial 30\% enrichment only costs one-fifth of the full effort and could enable earlier physics data taking with fractional effort.

To understand the impact on the projected sensitivity of different enrichment strategies, we plot the projected 90\% \ac{CL} exclusion limit for the \ac{0nu2b}-decay in Figure~\ref{fig:enrichment_sensitivity} as a function of enrichment of the inner hermetic volume. At 22\% enrichment HERETIX will match the projected sensitivities of the XLZD 60\,tonne nominal design proposal. Enrichment in excess of 50\% would be required to maximise the scientific reach of the hermetic \ac{TPC} and exceed the projected sensitivities of the nEXO experiment. The trade-off between minimum total effort and earlier physics deliverable must therefore be made on the basis of the scientific goal and resources of the experiment. 

\subsection{Cryogenic design, purification and operation}\label{sec:xenon_handling}
The hermetic separation and cryogenic design of the sapphire cylinder presents a significant challenge. Hermetic \ac{TPC} designs are typically proposed to reduce the radon-induced background from surface contamination of the detector materials, demanding a low leak-rate into the radon-free inner volume. In this work, the main motivation for the hermetic volume is driven by the containment of the \isotope[136]{Xe} enriched mass of the inner volume. It is crucial to ensure a stable concentration over the lifetime of the experiment. The choice of the wall material for the inner volume dictates the sealing type and design. We aim for a fully hermetic bond of the inner vessel structure, with brazing of thin metallic layers to seal the joints. However, in case mechanical and technical constraints require, the sealing should ensure that the \isotope[136]{Xe} concentration stays within 1\% over the lifetime of the experiment and therefore should be smaller than $10^{-2}$\,mbar$\cdot$L/s. For sapphire, a standard vacuum-sealing method with low leak rates can be selected to meet the requirement. Similarly, cryogenic operations will be performed completely separately, so that no mixing of the inner and outer volumes occurs.

The typically high thermal conductivity of the inner vessel material allows for cooling and liquefaction of the inner volume by the outer volume. However, a fast acting active cooling unit like a pulse tube refrigerator is needed to compensate for additional heat influxes and enable a stable xenon pressure and temperature. A passive pressure compensation device that equilibrates short-term differences is foreseen  to ensure the integrity of the precious content of the inner vessel. The placement of the inner volume within the field cage of the outer \ac{TPC} prohibits the use of conventional sensors such as capacitive levelmeters, but the liquid level could be measured using differential pressure sensors utilising the gas lines.

The hermetic separation requires an independent gas purification and recovery system for the enriched xenon volume. Electronegative impurities could be removed by standard gas getters using radon-free gas pumps\cite{Schulte_2021,XENON:2024wpa}, which would also enable the use of typical calibration sources for the inner volume. These calibration sources can also be used to estimate the leak tightness of the inner vessel. The removal of radioactive impurities such as \isotope[85]{Kr} and the remaining \isotope[222]{Rn} would be performed by online distillation~\cite{XENON1TOnlineDST} using a low flow distillation column which can be switched between both operation modes. However, these radioactive impurities do not influence the \ac{WIMP} or \ac{0nu2b} sensitivity presented in this work. Beyond purification, the gas system should also include an option for periodic measurement of the \isotope[136]{Xe} concentration.

\section{Discussion}

In this work we presented HERETIX, a proposal for a unified rare-event observatory for \ac{DM} and \ac{0nu2b} searches using \ac{LXe} \acp{TPC} as the common detector technology. Our proposal envisions a dual-phase \ac{TPC} divided into two volumes hermetically sealed from each other, with the inner and outer volumes containing enriched and depleted \isotope[136]{Xe}, optimised for \ac{0nu2b} and \ac{DM} searches, respectively. This construction allows a significant reduction in the amount of xenon compared to running the two searches in separate detectors, while also enabling shared detector infrastructure. Because only the inner volume is enriched, far less natural-abundance xenon is needed as input to the enrichment process than would be required to enrich the entire XLZD target to competitive \isotope[136]{Xe} concentrations. The main motivation is that the enriched inner volume benefits from the shielding provided by the depleted outer one, while the outer volume profits from the reduced \ac{2nu2b}-induced ER background.

The practical challenges of simultaneously separating the two volumes and maintaining access to them for recirculation and cooling require further research and development. The presence of the inner volume alleviates the requirement for multi-meter scale wire-meshes, and prompted the HERETIX design of a solid cathode and anode with a deposited mesh. HERETIX relies on sapphire for the material of the inner vessel and solid electrodes, but several properties need to be fully studied in \ac{LXe} to realize the proposed detector. The absorption of the xenon scintillation 178\,nm VUV light is not known, and is assumed to be on par with quartz for HERETIX sensitivity presented in this work. With regards to manufacturing, the processes of fusing individual sapphire pieces together, and providing electrical connectivity to the gate electrode (which remains a wire-mesh) must be validated and tested for \ac{LXe} \acp{TPC}.

The projected sensitivity to $T_{1/2}^{0\nu\beta\beta}$ of HERETIX reaches a world-leading $2.5 \times 10^{28} \, \mathrm{years}$ for a $10 \, \mathrm{year}$ exposure, with a \ac{WIMP} sensitivity comparable to that of XLZD. In the more optimistic scenario, where \isotope[238]{U} activity in the \acp{PMT} is reduced by 90\% from current screening values and sapphire with < 1\,$\upmu$Bq/kg of \isotope[238]{U} is procured for construction of the inner vessel, HERETIX offers a virtually background free search, despite this work assuming cryostat activities on par with that of the XENONnT experiment. HERETIX, in this scenario, would observe less than 2 events from material backgrounds in the \ac{RoI} during a 10\,year exposure, resulting in a projected sensitivity of $3.2 \times 10^{28} \, \mathrm{years}$. Implementation of a more nEXO-like scenario, in which the inner volume is operated as a single-phase \ac{TPC} with charge readout, thus improving the energy resolution around $Q_{\beta\beta}$ and the ability to discriminate the \ac{2nu2b} background would then be advisable to extend the sensitivity even further.
 
\begin{acknowledgements}
This work was supported by Pazy foundation, ISF grant 1859/22 and the European Research Council (ERC) through the ERC AdG project “LowRad”. This project has received funding from the European Research Council under the European Union’s Horizon 2020 research and innovation program (Grant agreement No. 101055063).
\end{acknowledgements}

\bibliographystyle{JHEP}
\bibliography{biblio.bib}

\end{document}